\documentclass[superscriptaddress, twocolumn, prl]{revtex4-1}

\usepackage{graphicx}
\usepackage{xcolor}
\usepackage{physics}

\begin{document}
\title{Fault-tolerant error correction with the gauge color code}

\author{Benjamin J. Brown}
\affiliation{Niels Bohr International Academy, Niels Bohr Institute, Blegdamsvej 17, 2100 Copenhagen, Denmark}
\affiliation{Quantum Optics and Laser Science, Blackett Laboratory, Imperial College London, Prince Consort Road, London SW7 2AZ, United Kingdom}
\author{Naomi H. Nickerson}
\affiliation{Quantum Optics and Laser Science, Blackett Laboratory, Imperial College London, Prince Consort Road, London SW7 2AZ, United Kingdom}
\author{Dan E. Browne}
\email{d.browne@ucl.ac.uk}
\affiliation{Department of Physics and Astronomy, University College London, Gower Street, London WC1E 6BT, United Kingdom}

\begin{abstract}

The constituent parts of a quantum computer are inherently vulnerable to errors. To this end we have developed quantum error-correcting codes to protect quantum information from noise. However, discovering codes that are capable of a universal set of computational operations with the minimal cost in quantum resources remains an important and ongoing challenge. One proposal of significant recent interest is the gauge color code. Notably, this code may offer a reduced resource cost over other well-studied fault-tolerant architectures using a new method, known as gauge fixing, for performing the non-Clifford logical operations that are essential for universal quantum computation. Here we examine the gauge color code when it is subject to noise. Specifically we make use of single-shot error correction to develop a simple decoding algorithm for the gauge color code, and we numerically analyse its performance. Remarkably, we find threshold error rates comparable to those of other leading proposals. Our results thus provide encouraging preliminary data of a comparative study between the gauge color code and other promising computational architectures.

\end{abstract}

\maketitle

\section{Introduction}

Scalable quantum technologies require the ability to maintain and manipulate coherent quantum states over an arbitrarily long period of time. It is problematic then that the small quantum systems that we might use to realize such technologies decohere rapidly due to unavoidable interactions with the environment. To resolve this issue we have discovered quantum error-correcting codes~\cite{Shor95, Steane96} which make use of a redundancy of physical qubits to maintain encoded quantum states with arbitrarily high fidelity over an indefinite period.

Ideally, we will design a fault-tolerant quantum computer that requires as few physical qubits as possible to minimize the resource cost of a quantum processor, and indeed, the cost in resources of a computational architecture is very sensitive to the choice of quantum error-correcting code used by a fault-tolerant scheme. It is therefore of great interest to analyse different quantum error-correction proposals to compare and contrast their resource demands.

Color codes~\cite{Bombin06, Bombin07, Bombin10, Bombin13, Bombin13a} are a family of topological quantum error-correcting codes~\cite{Kitaev03, Dennis02, Terhal13, LidarBrun} with impressive versatility~\cite{Eastin, BravyiKonig13, Kubica15} for performing fault-tolerant logic gates~\cite{Bombin11, Fowler11, Landahl14}. This is an important consideration as we search for schemes that realise fault-tolerant quantum computation with a low cost in quantum resources. In particular, a fault-tolerant quantum computer must be able to perform a non-Clifford operation, such as the $\pi / 8$-gate, to realise universal quantum computation.

In general, performing Non-Clifford gates can present a considerable resource cost over the duration of a quantum computation. As such, the resource cost of realising scalable quantum computation is sensitive to the method a fault-tolerant computational scheme uses to realise non-Clifford gates. To this end the gauge color code~\cite{Bombin13a, Kubica14, Bombin14a, Bombin14b} has attracted significant recent interest because, notably, this three-dimensional quantum error-correcting code can achieve a universal gate set via gauge fixing~\cite{Paetznick13, Anderson14}.

In contrast to the gauge-color code, surface code quantum computation, a leading approach towards low-resource quantum computation~\cite{Kitaev03, Dennis02, Raussendorf07}, makes use of magic state-distillation~\cite{BravyiKitaev05} to perform $\pi/8$-gates. Magic state distillation can be achieved with $\mathcal{O}(L^3)$ space-time resource cost~\cite{Raussendorf07, Fowler12a, Bravyi12, Meier13, Li14}. Similarly, the gauge color code performs $\pi/8$-gates via gauge fixing in constant time~\cite{Bombin14a}, and as such has an equivalent scaling in space-time resource cost as the surface code since the gauge color code requires $\mathcal{O}(L^3)$ physical qubits. However, given that gauge fixing requires no additional offline quantum resources to perform a non-Clifford rotation, the gauge color code may reduce the quantum resources that are necessary for fault-tolerant quantum computation by a constant fraction.

It is also noteworthy that the gauge color code is local only in three dimensions and as such, unlike the surface code, cannot be realised using a two-dimensional array of locally interacting qubits. Instead, the gauge color code may be an attractive model for non-local quantum-computational architectures such as networked schemes~\cite{Barrett05, Fujii12, Nickerson13, Monroe14, Nickerson14}.

Given the significant qualitative differences between the gauge color code and the surface code, it is interesting to perform a comparative analysis of these two proposals. In this Manuscript we investigate error correction with the gauge color code.

Dealing with errors that continually occur on physical qubits is particularly difficult in the realistic setting where syndrome measurements can fail and return false readings~\cite{Dennis02}. Attempting to correct errors using inaccurate syndrome information will introduce new physical errors to the code. However, given enough syndrome information, we can distinguish measurement errors from physical errors with enough confidence that the errors we introduce are few, and can be identified at a later round of error correction~\cite{Bombin14a}. In the case of the toric code~\cite{Dennis02}, we accumulate sufficient error data by performing multiple rounds of syndrome measurements. Surprisingly, the structure of the gauge color code enables the acquisition of fault-tolerant syndrome data using only one round of local measurements~\cite{Bombin14a}. This capability is known as single-shot error correction.

Here, we obtain a noise threshold for the gauge color code using a phenomenological noise model where both physical errors and measurement faults occur at rate $p$. We develop a single-shot decoder to identify the sustainable operating conditions of the code, i.e. the noise rate below which information can be maintained arbitrarily well, even after many cycles of error correction. We estimate a sustainable error rate of $p_{\text{sus}} \sim 0.31\%$  using an efficient clustering decoding scheme~\cite{Harrington, BravyiHaah11A, Anwar14, Hutter14, Watson14, Wootton15} that runs in time $\mathcal{O}(L^6 \log L)$,~\cite{BravyiHaah11A}, where the distance of the code is $d = L + 2$,~\cite{Bombin13a}. Remarkably, the threshold we obtain falls within an order or magnitude of the optimal threshold for the toric code under the same error model, $\sim2.9\%$,~\cite{Wang03}. Furthermore, we also use our decoder to estimate how the logical failure rate of the gauge color code scales below the threshold error rate by fitting to a heuristic scaling hypothesis.

\section{Results}

\begin{figure}
\includegraphics{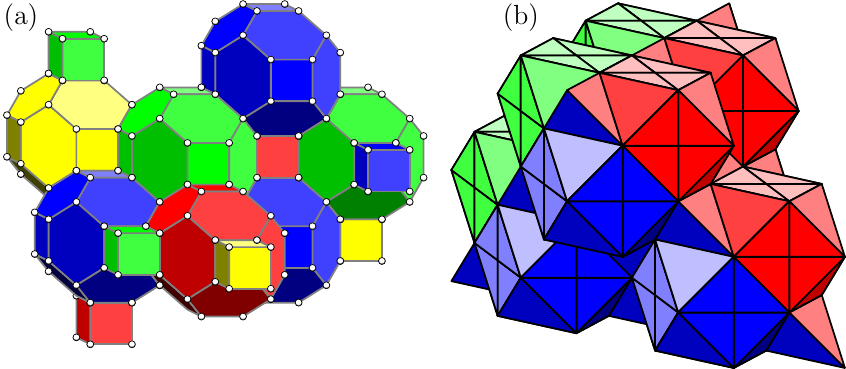}
\caption{(a)~In the primal picture qubits sit on the vertices of a four-valent lattice. The three-dimensional cells of the lattice are four colorable, i.e., every cell can be assigned one of four colors such that it touches no other cell of the same color. (b)~The gauge color code of linear size $L=5$ drawn in the dual picture. Qubits lie on simplices of the lattice. The faces of the tetrahedra in the figure are given one of four colors such that no two faces of a given tetrahedron have the same color. The tetrahedra are stacked such that faces that touch always have the same color. For the gauge color code to encode a single logical qubit, the lattice must have four distinct, uniformly colored boundaries, as is shown in the Figure. We give more details on the lattice construction in the dual picture in Supplementary Note~1 together with Supplementary Figures~1-6.  \label{ColorCodeLattice}}
\end{figure}

\subsection{The gauge color code}

The gauge color code is a subsystem code~\cite{Poulin05} specified by its gauge group, $\mathcal{G}$. From the center of the gauge group, $Z(\mathcal{G})$, we obtain the stabilizer group for the code, $\mathcal{S} = Z(\mathcal{G}) \cap \mathcal{G}$, and its logical operators, $\mathcal{L} = Z(\mathcal{G}) \backslash \mathcal{G}$. Elements of the stabilizer group, $S \in \mathcal{S}$, satisfy the property that $S | \psi \rangle = | \psi \rangle$ for all codewords of the code $| \psi \rangle$.

The code is defined on a three-dimensional four-valent lattice of linear size $L$ with qubits on its vertices~\cite{Bombin07, Bombin07a}. The lattice must also be four-colorable, i.e. each cell of the lattice can be given a color, red, green, yellow, or blue, denoted by elements of the set  $\mathcal{C} = \{ \mathbf{r}, \mathbf{g}, \mathbf{y}, \mathbf{b} \} $, such that no two adjacent cells are of the same color. The lattice we consider is shown in Fig.~\ref{ColorCodeLattice}(a), where the cells are the solid colored objects in the Figure.

The cells of the lattice define stabilizer generators for the code, while the faces of cells define its gauge generators. The gauge operators, otherwise known as face operators, are measured to infer the values of stabilizer operators. The face operators have weight four, six and eight, where the weight-eight face operators lie on the boundary of the lattice. More specifically, for each cell $c$ there are two stabilizer generators, $S^X_c = \prod_{j\in \mathcal{V}(c)}X_j$ and $S^Z_c = \prod_{j\in \mathcal{V}(c)} Z_j$, where $\mathcal{V}(c)$ are the qubits on the boundary vertices of cell $c$, and $X_j$ and $Z_j$ are Pauli-X and Pauli-Z operators acting on vertex $j$. For each face $f$ there are two face operators, $G_f^X = \prod_{j\in \mathcal{V}(f)} X_j$ and $G_f^Z= \prod_{j\in \mathcal{V}(f)} Z_j$ where $\mathcal{V}(f)$ are the vertices on the boundary of face $f$. We will call the outcome of a face operator measurement a face outcome.

Given suitable boundary conditions~\cite{Bombin07a, Bombin13a}, the code encodes one qubit, whose logical operators are $\overline{X}  = \prod_{j \in \mathcal{Q}} X_j$ and $\overline{Z}  = \prod_{j \in \mathcal{Q}} Z_j$, where $\mathcal{Q}$ is the set of physical qubits of the code. A lattice with correct boundaries is conveniently represented on a dual lattice of tetrahedra using the convention given in Ref.~\cite{Bombin13a}. The lattice we consider is illustrated in Fig.~\ref{ColorCodeLattice}(b), where we discuss its construction in detail in the Supplementary Notes~1 and~2, together with Supplementary Figures~1-7. Importantly, we require that the lattice has four boundaries, distinguished by colors from set $\mathcal{C}$.

\subsection{Single-shot error correction}

\begin{figure}[b]
\includegraphics{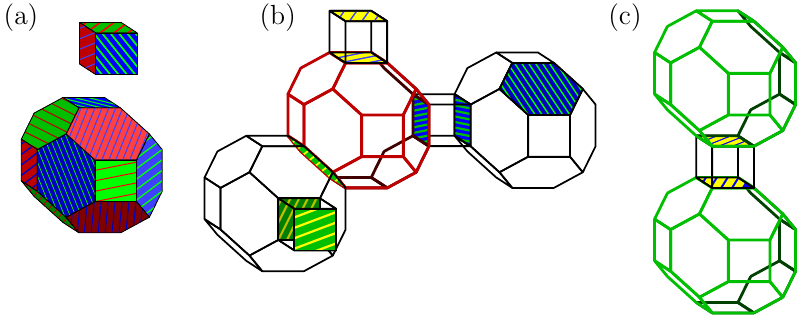}
\caption{(a)~The faces of the cells of the color code are three colorable. (b)~An example of a set of face measurement outcomes where measurements are reliable. Face operators that return value -1 are colored, otherwise they are left transparent. The stabilizer at the cell with thick red edges contains a stabilizer defect. This cell has one face operator of each color returning a -1 outcome. All other cells have an even parity of -1 face  outcomes over their colored subsets,   indicating that these cells contain no stabilizer defect. (c)~A gauge syndrome where colored subsets of face measurements about the green cells do not agree, thus indicating measurement errors. \label{GaugeMeasurements}}
\end{figure}

A quantum error-correcting code is designed to identify and correct errors. Due to the symmetry of the gauge group, it suffices here to consider only bit-flip, i.e. Pauli-X errors. We consider a phenomenological error model consisting of physical errors and measurement errors. A physical error is a Pauli-X error on a qubit, while a measurement error returns the opposite outcome of the correct reading. Errors will be identically and independently generated with the same probability $p$.

In a stabilizer code, errors are identified by stabilizer measurements that return eigenvalue $-1$, which we call stabilizer defects. We use the stabilizer syndrome, a list of the positions of stabilizer defects, to predict the incident error. In the gauge color code we do not measure stabilizer operators directly, but instead infer their values by measuring face operators, which is possible due to the fact that $\mathcal{S} \subseteq \mathcal{G}$.

In addition to using face outcomes to infer stabilizer eigenvalues, we can also exploit the local constraints in $\mathcal{G}$ of the gauge color code to detect and account for measurement errors. Remarkably, measurement errors can be detected reliably by measuring each face operator only once,  so called single-shot error correction \cite{Bombin14a}.

The local constraints stem from the structure of the code. We first observe that the faces of a cell are necessarily three colorable, as shown in Fig.~\ref{GaugeMeasurements}(a). It follows from the three colorability of the faces that the product of the gauge operators $G^Z_f$ of any of the differently colored subsets of faces of cell $c$ recover the stabilizer operator $S^Z_c$. By measuring all the faces of the lattice, we redundantly recover each stabilizer eigenvalue three times where the three outcomes of a given cell are constrained to agree. In Fig.~\ref{GaugeMeasurements}(b) we show an example of a gauge measurement configuration where the outcomes are reliable. Following this observation, we can use violations of the local constraints about a cell to indicate the positions of measurement errors. In Fig.~\ref{GaugeMeasurements}(c) we show a syndrome where the product ofthe face outcomes of different subsets of of two cells, colored with thick green edges, do not agree.

Fault-tolerant decoding with the gauge color code proceeds in two stages. The first stage, syndrome estimation, uses face outcomes that may be unreliable to estimate the locations of stabilizer defects. The second stage, stabilizer decoding, takes the estimated stabilizer defect locations and predicts a correction operator to reverse physical errors.

\begin{figure}
\includegraphics{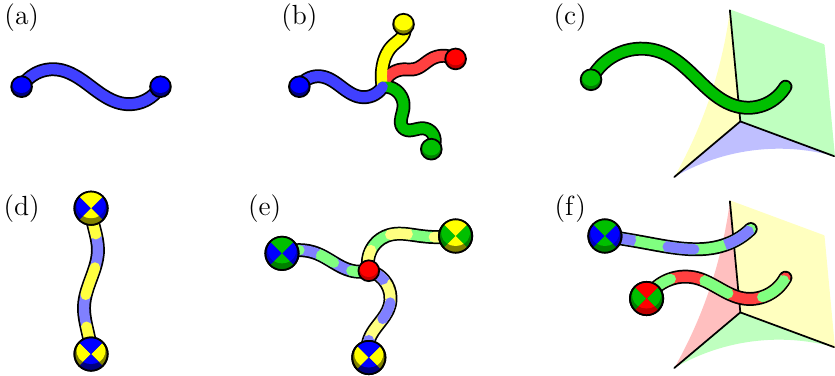}
\caption{\label{Syndrome}(a) Two blue syndromes indicate the end points of a string of errors connecting the two points. (b) A string error branches to create one syndrome of each color. (c) A green string error can terminate at a green boundary, thus generating only a single green syndrome. 
(d)~A pair of $\mathbf{yb}$ gauge defects, shown by the vertices, can be caused by a string of incorrect $\mathbf{yb}$ face outcomes on the dual lattice. The displayed gauge syndrome is equivalent to that shown in Fig.~\ref{GaugeMeasurements}(c) where the measurement errors have occurred on the faces that returned $ - 1 $ outcomes. (e)~Three gauge defects, colored $\mathbf{gy}$, $\mathbf{gb}$ and $\mathbf{yb} $, can be caused by an error string that branches at a red cell. The branching point indicates a stabilizer defect at a red cell. (f)~Strings of incorrect face outcomes of colors $\mathbf{rg}$ and $\mathbf{gb}$ terminate at a yellow boundary.}
\end{figure}

The topological nature of the gauge color code is such that its stabilizer defects satisfy conservation laws that enable us to employ well-studied decoding algorithms~\cite{Duclos-CianciPoulin10, BravyiHaah11A, WoottonLoss12} to complete stabilizer decoding. The syndrome identified by stabilizer measurements is studied extensively in Refs.~\cite{Bombin07, Bombin07a}. Stabilizers that return $-1$ outcomes occur in pairs at the endpoints of strings of errors, whose endpoints take the color of its terminal cells, as shown in Fig.~\ref{Syndrome}(a). As such, one can also regard an error string as carrying the color of the stabilizer defects at its endpoints. Error strings on the gauge color code can also branch into three strings of conjugate colors, thus creating one syndrome of each color, as shown in Fig.~\ref{Syndrome}(b). Finally, error strings of a given color can terminate at the boundary of their respective color, as shown in Fig.~\ref{Syndrome}(c). In our simulation we decode stabilizer defects by adapting a clustering decoder~\cite{Harrington, BravyiHaah11A, Anwar14, Hutter14, Watson14, Wootton15} where clusters grow linearly.

We concentrate now on syndrome estimation. Syndrome estimation uses a gauge syndrome, a list of gauge defects, to estimate a stabilizer sydrome. A lattice cell can contain as many as three gauge defects. Gauge defects are distinguished by a color pair $\mathbf{uv}$, with $\mathbf{u},\, \mathbf{v} \in \mathcal{C}$, such that $\mathbf{u} \not= \mathbf{v}$ and $\mathbf{uv} = \mathbf{vu}$. The color pair of a gauge defect relates to the coloring of the lattice faces. A face is given the color pair opposite to the colors of its adjacent cells, i.e. the face shared by two cells with colors $\mathbf{r}$ and $\mathbf{g}$ is colored $\mathbf{yb}$. A cell $c$ contains a $\mathbf{uv}$ gauge defect if the product of all the $\mathbf{uv}$ face outcomes bounding $c$ is $-1$. Following this definition, a stabilizer defect is equivalent to three distinct gauge defects in a common cell.

Studying the gauge syndrome enables the identification of measurement errors. We consider face $f$, colored $\mathbf{uv}$, that is adjacent to cell $c$. In the noiseless measurement case, where $c$ contains no stabilizer defect, by definition, cell $c$ should contain no gauge defects. However, if face $f$ returns an incorrect outcome, we identify a $\mathbf{uv}$ gauge defect at $c$. Conversely, if cell $c$ contains a stabilizer defect in the ideal case, and $f$ returns an incorrect outcome, no $\mathbf{uv}$ gauge defect will appear in $c$. With these examples we see that cells that contain either one or two gauge defects indicate incorrect face outcomes.

An incorrect face outcome affects gauge defects in both of its adjacent cells. In general, incorrect face outcomes of color $\mathbf{uv}$ form error strings on the dual lattice, whose end points are $\mathbf{uv}$ gauge defects, where individual incorrect face outcomes are segments of the string. Error strings of incorrect $\mathbf{uv}$ face outcomes changes the parity of $\mathbf{uv}$ gauge defects at both of its terminal cells.

We require an algorithm that can use gauge syndrome data to predict a likely measurement error configuration, and thus estimate the stabilizer syndrome. We adapt the clustering decoder~\cite{Harrington, BravyiHaah11A, Anwar14, Hutter14,  Watson14} for this purpose. The decoder combines nearby defects into clusters that can be contained within a small box. Clusters increase linearly in size to contain other nearby defects until they contain a set of gauge defects that can be caused by a measurement error contained within the box. Once clustering is completed, a correction supported inside the boxes is returned.

We briefly elaborate on correctable configurations of gauge defects. Pairs of $\mathbf{uv}$ gauge defects are caused by strings of incorrect $\mathbf{uv}$ face outcomes, and therefore form correctable configurations, as shown in Fig.~\ref{Syndrome}(d). As an example, the gauge syndrome in Fig.~\ref{Syndrome}(d) depicts the gauge defects and measurement error string shown in Fig.~\ref{GaugeMeasurements}(c) where the measurement errors have occurred on the faces that returned $-1$ measurement outcomes. Triplets of gauge defects, colored $\mathbf{uv}$, $\mathbf{vw}$ and $\mathbf{uw}$, also form correctable configurations. Error strings that cause triplets of correctable gauge defects branch at a cell, and thus indicate a stabilizer defect, shown in Fig.~\ref{Syndrome}(e). The stabilizer defect where the error string branches lies at a cell colored $\mathbf{x} \not= \mathbf{u}, \mathbf{v}, \mathbf{w} $. Gauge defects can also arise due to incorrect face outcomes on the lattice boundary. Specifically, the boundary colored $\mathbf{w}$ contains faces of color $\mathbf{uv}$ where $ \mathbf{u}, \mathbf{v} \not= \mathbf{w}$. With this coloring we can find correctable configurations of single $\mathbf{uv}$ gauge defects, together with a boundary of color $\mathbf{w}$, see Fig.~\ref{Syndrome}(f). In general, a cluster can contain many correctable pairs and triplets of gauge defects.

As we have mentioned, correctable clusters of gauge defects can give rise to stabilizer defects. It is important to note that the error-correction procedure is sensitive to the positions of stabilizer defects within a correctable cluster, as discrepancies in their positions later affect the performance of the stabilizer decoding algorithm. As such we must place stabilizer defects carefully. For cases where a correctable cluster of gauge defects returns stabilizer defects, we assign their positions such that they lie at the mean position of all the gauge defects within the correctable cluster, at the nearest cell of the appropriate color. Once syndrome estimation is complete, the predicted stabilizer syndrome is passed to the stabilizer decoder, and a correction operator is evaluated.

We remark that gauge defects can be incorrectly analyzed during syndrome estimation. In which case, measurement errors sometimes masquerade as stabilizer defects, and sometimes stabilizer defects can be misplaced.  We will then attempt to decode the incorrect stabilizer syndrome and mistakenly introduce errors to the code. In general, any error-correction scheme that takes noisy measurement data will introduce residual physical errors to a code. These errors can be corrected in the future, provided the remaining noise is of a form that a decoder can correct. In general however, one must worry that large correlated errors can be introduced that adversarially corrupt encoded information~\cite{Aharonov06, Ng09, Preskill13, Jouzdani14, Fowler14, Hutter14a}. Such errors may occur in the gauge color code if, for instance, we mistakenly predict two stabilizer defects of the same color that are separated by a large distance. We give an example of a mechanism that might cause a correlated error during syndrome estimation with the gauge color code in the Supplementary Note~3, together with Supplementary Figure~8.

A special property of the gauge color code is that measurement errors, followed by syndrome estimation, will only introduce false defects in locally correctable configurations. Therefore, residual errors remain local to the measurement error. Moreover, the code is such that the probability of obtaining  configurations of face outcomes that correspond to faux stabilizer defects decays exponentially with the separation of their cells. This is because the number of measurement errors that must occur to produce a pair of false stabilizer defects is extensive with their cell separation. To this end, the errors introduced from incorrect measurements are local to the measurement error and typically small. This property, coined `confinement' in Ref.~\cite{Bombin14a}, is essential for fault-tolerant error correction. Most known codes achieve confinement by performing syndrome measurements many times. We give numerical evidence showing that our error-correction protocol confines errors in the following Subsection.

\subsection{The Simulation}

We simulate fault-tolerant error correction with encoded states $|\psi_j \rangle $ of linear size $L$ where $j$ indicates the number of error-correction cycles that have been performed, and where $|\psi_0 \rangle$ is a codeword. We seek to find a correction operator $C$ such that $CE \in \mathcal{G}$ where $E$ is the noise incident to $|\psi_0 \rangle$ after $N$ error-correction cycles.

To maintain the encoded information over long durations, we repeatedly apply error-correction cycles to keep the physical noise sufficiently benign. After a short period, the state $|\psi_{j-1} \rangle$ will accumulate physical noise $E_{j}(p) $ with error rate $p$. To correct the noise, we first estimate a stabilizer syndrome, $\mathbf{s}_{j}$, using gauge syndrome data with the syndrome estimation algorithm $M_q$, where measurement outcomes are incorrect with probability $q = p$. Specifically, we have that $\mathbf{s}_{j} = M_p(E_{j}(p) |\psi_{j-1} \rangle)$. We then use the stabilizer decoding algorithm $D$ to predict a suitable correction operator $C_{j}= D(\mathbf{s}_{j})$, such that we obtain
\begin{equation}
|\psi_{j} \rangle =C_{j} E_{j}(p) |\psi_{j-1} \rangle.
\end{equation}

It is important to note is that $ |\psi_{j} \rangle $ is not necessarily in the code subspace. For $q>0$, stabilizer syndromes will in general be incorrectly estimated, and thus the correction operator $C_{j}$ introduces some new errors to the code. 

\begin{figure}
\includegraphics[width=\columnwidth]{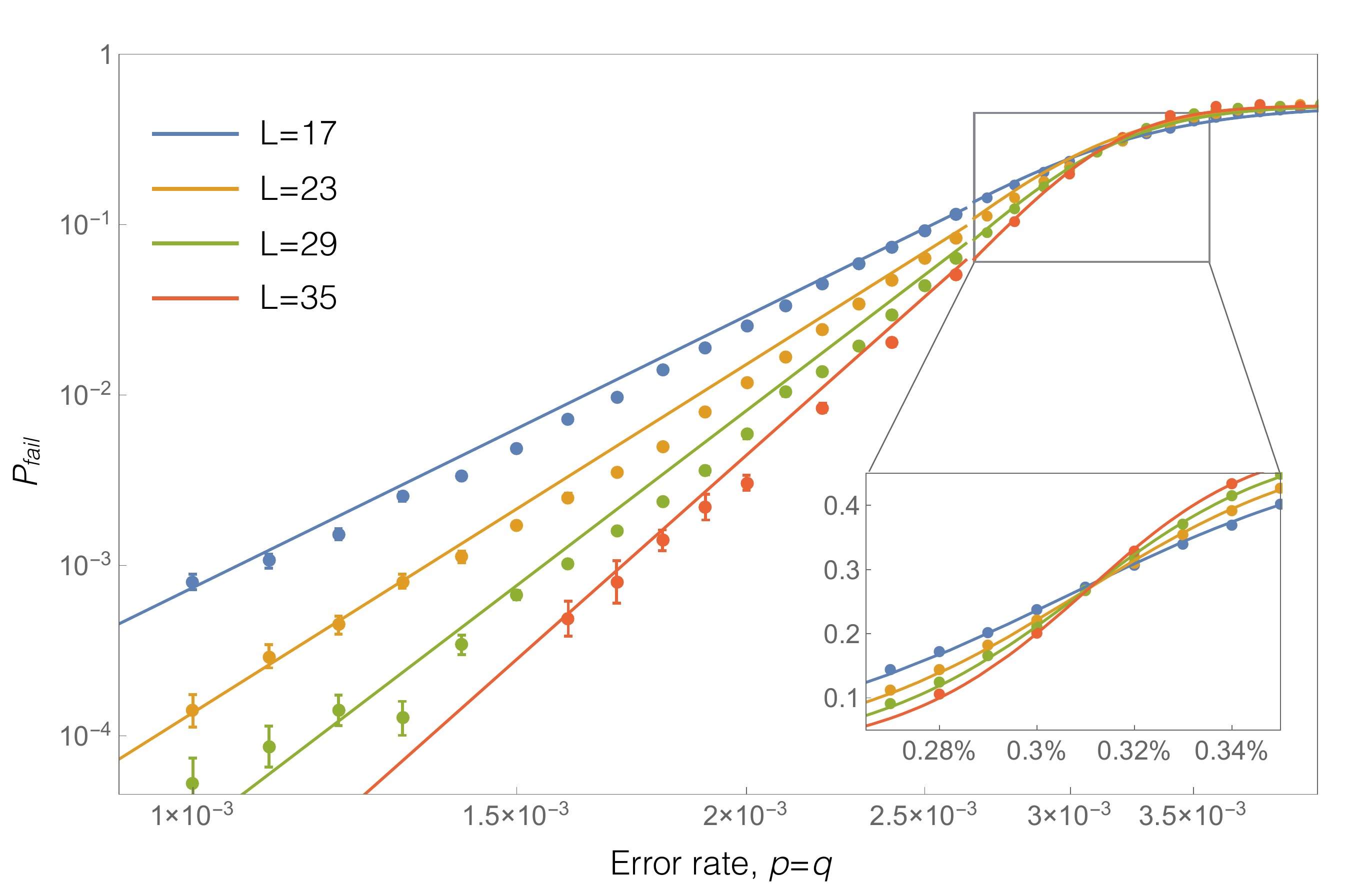}
\caption{\label{ThresholdFigure}We show logical error rates, $P_{\text{fail}}$, as a function of physical error rate, $p$, for system sizes $L = 17,\, 23,\, 29 $ and $35$ shown in blue, yellow, green and red, respectively, as is marked in the legend, where we collect data after $N = 8$ rounds of error correction during which measurements are performed unreliably. The error bars show the standard error of the mean given by the expression~$\Delta p =  \sqrt{ (1 - P_{\text{fail}})P_{\text{fail}} /\eta } $ where $\eta$ is the number of Monte Carlo samples we collect.  The data used to determine the threshold error rate is shown in the inset, where we determine the threshold using the fitting described in the Methods Section. The fitting is also plotted in the inset. In the main Figure the solid lines show the fitted expression, Eqn.~(\ref{Eqn:Low-pFitting}), to demonstrate the agreement of our scaling hypothesis with numerically evaluated logical error rates where $p < p_{\text{th}}$. We remark that the fitting is made using data for all values of $N$, and not only the data shown in this plot, as we explain in the Methods Section.}
\end{figure}

We require that after $N$ error-correction cycles we can estimate error $E$ of state $ |\psi_{N} \rangle = E | \psi_0 \rangle $ to perform a logical measurement. To perform the $\overline{Z}$ logical measurement~\cite{Alicki10, Bombin14a}, we measure each individual qubit of the code in the Pauli-Z basis. This destructive transversal measurement gives us the eigenvalues of stabilizers $S_c^Z$ to diagnose $E$, and to thus recover the eigenvalue of $\overline{Z}$.

During readout, measurement errors and physical errors  have an equivalent effect; both appear as bit flips. To simulate errors that occur during the readout process, we apply the noise operator $E(p)$ to the encoded state before decoding. We therefore calculate logical failure rates
\begin{equation}
P_{\text{fail}}(N)  = \text{prob}(C(N)E(p) E  \in \mathcal{G} ), \label{Eqn:LogicalFailureCalculation}
\end{equation}
where $C(N) = D(M_0(E(p)| \psi_N \rangle))$. We evaluate $P_{\text{fail}}(N)$ values using Monte Carlo simulations.

To analyze the performance of the proposed decoding scheme, we first look to find the sustainable error rate of the code, $p_{\text{sus}}$, below which we can maintain quantum information for an arbitrary number of correction cycles. The discovery of such a point suggests that the error-correlations caused by our correction protocol do not extend beyond a constant, finite, and decodable length, thus showing that we can preserve information indefinitely with arbitrarily high fidelity in the $ p < p_{\text{sus}}$ regime.

We define $p_{\text{sus}}$ as the threshold error rate, $p_{\text{th}}$, at the $N \rightarrow \infty $ limit, where the threshold is the error rate below which we can decrease $P_{\text{fail}}$ arbitrarily by increasing $L$~\cite{Terhal13}. In the inset of Fig.~\ref{ThresholdFigure} we show the near threshold data we use to evaluate a threshold, where we show the data for $N = 8$ as an example. We give the details of the fitting model we use to evaluate thresholds in the Methods Section of this Manuscript.

\begin{figure}
\includegraphics[width=\columnwidth]{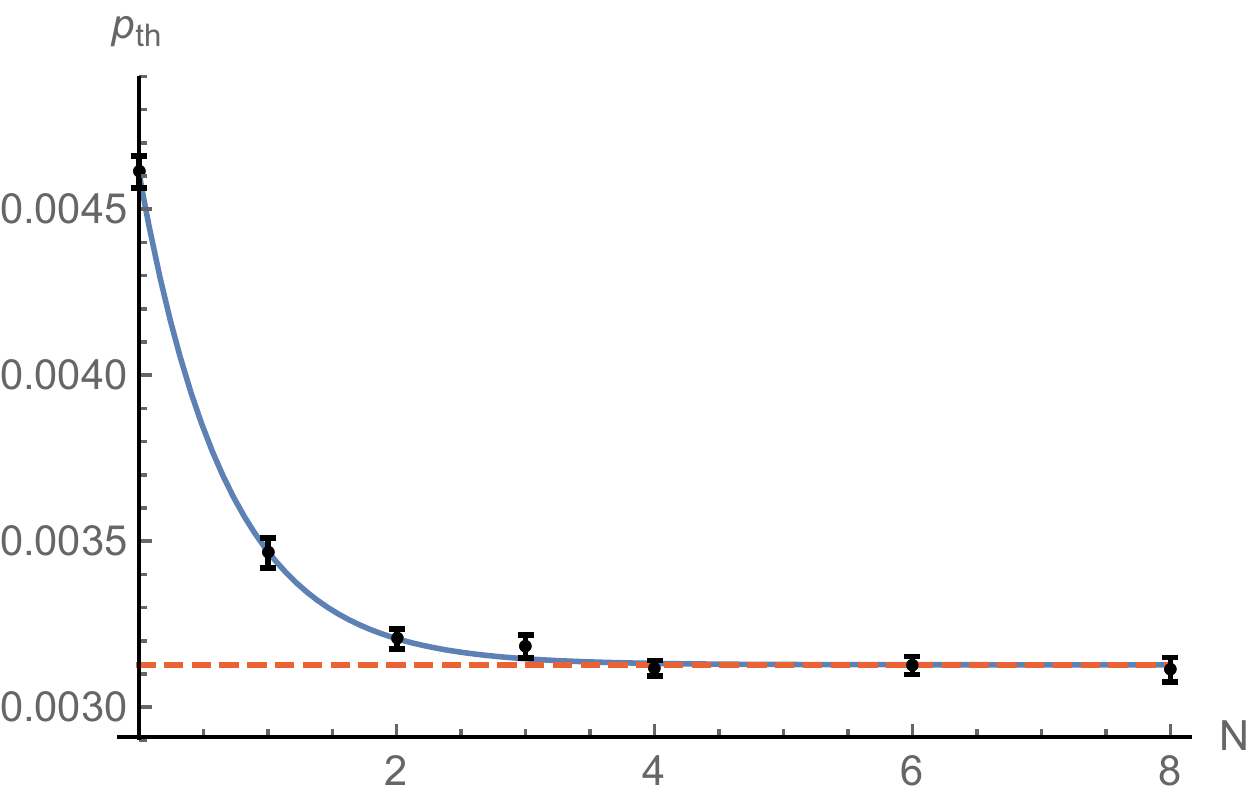}
\caption{\label{PhaseDiagram}Threshold error rates, $p_{\text{th}}$, are calculated with system sizes $L=23,\,29,\,35$ after $N$ error-correction cycles using $\eta \sim 10^4$ Monte Carlo samples. Error bars show the standard error of the mean which are determined using the {\bf NonLinearModelFit} function in {\bf Mathematica}. The solid blue line shows the fitting given in Eqn.~(\ref{Fitting}). The dashed red line marks $p_{\text{sus}} \sim 0.31 \%$, the sustainable noise rate of the code, the limiting value of $p_{\text{th}}$ from to the fitting as $N\rightarrow \infty$.}
\end{figure}

We next study the evaluated threshold values as a function of $N$. We show this data in the plot given in Fig.~\ref{PhaseDiagram}. The data shows that $p_{\text{th}}$ converges to $p_{\text{sus}} \sim 0.31\%$ where we fit for values $N \le 8$. We obtain this value with a fitting that converges to $p_{\text{sus}}$, namely
\begin{equation}
p_{\text{th}}(N) = p_{\text{sus}} \left[1 - (1 -  p_{\text{th}}(0) / p_{\text{sus}} ) \exp(-\gamma N)  \right]. \label{Fitting}
\end{equation}
We find $p_{\text{th}}(0) \sim 0.46 \%$ and $\gamma \sim 1.47 $. The convergent trend provides evidence that we achieve steady-state confinement in the high-$N$ limit, as is required of a practical error-correction scheme.

To verify further the threshold error rates we have determined, we next check that the logical failure rate decays as a function of system size in the regime where $p < p_{\text{th}}$. In the large $N$ limit, we fit our data to the following hypothesis
\begin{equation}
P_{\text{fail}}(N) = (N+1)A \exp \left(\alpha \log \left( \frac{p}{  p_{\text{sus}}} \right) d^\beta \right), \label{Eqn:Low-pFitting}
\end{equation}
where $A$, $\alpha$ and $ \beta $ are positive constants to be determined and $d$ is the distance of the code. We evaluate the variables in Eqn.~(\ref{Eqn:Low-pFitting}) as $A \sim 0.033$, $\alpha \sim 0.516$ and $\beta \sim 0.822$ using $\sim 5000$ CPU hours with data for $N \le 10$. Details on the fitting calculation are given in the Methods Section. We plot the fitted scaling hypothesis on Fig.~\ref{ThresholdFigure} for the case of $N = 8$ to show the agreement of Eqn.~(\ref{Eqn:Low-pFitting}) with the available data.

\section{Discussion}

To summarize, using only a simple decoding scheme, we have obtained threshold values that lie within an order of magnitude of the optimal threshold for the toric code under the phenomenological noise model. Moreover, we can expect that higher thresholds are achievable using more sophisticated decoding strategies~\cite{Duclos-CianciPoulin10, Wang10, Bombin12, WoottonLoss12, Sarvepalli12, HutterWoottonLoss13, Delfosse14, Anwar14, Herrold, Hutter14, Stephens14}. It may be possible to achieve a sufficiently high sustainable noise rate to become of practical interest, thus meriting comparison with the intensively-studied surface code~\cite{Fowler13, Fowler15}. To this end, further investigation is required to learn its experimental viability.

To continue such a comparative analysis, one should study the code using realistic noise models~\cite{Fowler12a} that respect the underlying code hardware. We expect that the threshold will suffer relative to the surface code when compared using a circuit-based noise model where high-weight gauge measurements are more error prone~\cite{Landahl11}. Fortunately, gauge color code lattices are known where face operators have weight no greater than six~\cite{Bombin13a}. While this is not as favorable as the weight-four stabilizer measurements of the surface code, given the ability to perform single-shot error correction, and $\pi/8$-gates through gauge fixing, we argue that the gauge color code is deserved of further comparison.


To the best of our knowledge, we have obtained the first threshold using single-shot error correction. Fundamentally, our favorable threshold is achieved using redundant syndrome data to identify measurement errors. It is interesting to ask if we can make use of a more intelligent collection of measurement data to improve thresholds further. Discovering single-shot error-correction protocols with simpler codes might help to address such questions.

\begin{acknowledgements}
We gratefully acknowledge S. Bartlett, H. Bomb\'{i}n, E. Campbell, S. Devitt,  A. Doherty, S. Flammia and M. Kastoryano for helpful discussions. Computational resources are provided by the Imperial College HPC Service. This work is supported by the EPSRC and the Villum Foundation. \\

\noindent {\bf Author contributions:} Original concept conceived by BJB and DEB, simulations designed and written by BJB and NHN, data collected and analysed by BJB and NHN, the manuscript was prepared by all the authors. \\

\noindent {\bf Data availability:} The code and data used in this Manuscript is available upon request to the corresponding author.\\

\noindent {\bf Competing financial interests:} The authors declare no competing financial interests.\\

\end{acknowledgements}

\section{Methods}

\subsection{Threshold calculations}
\label{SCTN:ThresholdCalculation}

The threshold error rate, $p_{\text{th}}$, is the physical error rate below which the logical failure rate of the code can be arbitrarily suppressed by increasing the code distance. We identify thresholds by plotting the logical failure rate as a function of physical error rate $p$ for several different system sizes, and identify the value $p = p_\text{th}$ such that $P_{\text{fail}}$ is invariant under changes in system size. In the main text we show the data used for a specific threshold calculation in Fig.~\ref{ThresholdFigure} where we use the $N = 8$ data as an example, and in Fig.~\ref{PhaseDiagram} we evaluate threshold error rates for the gauge color code as a function of the number of error correction cycles, $N$.

We evaluate threshold error rates by performing $\eta \sim 10^4$ Monte Carlo simulations for each value of $p$ close to the crossing point using codes of system sizes $L = 23, \, 29$ and $35$, except for the case that $N=0$ where we evaluate the logical failure rate with system sizes $L = 31,\,39$ and $47$. Simulating larger system sizes where $N = 0$ is possible as in this case we read out information immediately after encoding it, such that $E = 1$ as shown in Eqn.~(\ref{Eqn:LogicalFailureCalculation}). We therefore need not perform syndrome estimation in the $N= 0$ simulation. The threshold error rate at $N=0$ is thus the threshold error rate of the clustering decoder for the gauge color code where measurements are performed perfectly, i.e., $ q = 0$.

We identify the crossing point by fitting our data to the following formula
\begin{equation}
 P_{\text{fail}} = B_0 + B_1 x + B_2 x^2,
\end{equation}
where $x = (p - p_{\text{th}})L^{1/\mu}$ and $B_j$, $p_{\text{th}}$ and $\mu$ are constants to be determined. We show an example of this fitting in the Inset of Fig.~\ref{ThresholdFigure}.

At the threshold error rate the code produces logical failures at a rate between $0.075$ and $0.27$ depending on $N$. We expect such behavior as the number of logical failures will increase with repeated use of a decoder. We therefore obtain between $\sim 750$ and $\sim2700$ logical failures per data point close to the threshold error rate.

\subsection{Overhead analysis}
\label{SubSec:Overheads}

Here we summarize the resource-scaling analysis we give in the $p < p_{\text{th}}$ regime. We suppose the logical failure rate in this regime scales like
\begin{equation}
P_{\text{fail}}(N) \approx (N+1)A \exp( \alpha \log (p / p_{\text{th}}(N))   d^\beta  ), \label{Eqn:LogicalFailureRate}
\end{equation}
where  $A$, $\alpha$ and $\beta$ are constants to be determined, $d $ is the distance of the code, $p$ is the error rate and $N$ is the number of uses of the decoder we make before readout. The value of $p_{\text{th}}(N)$ is determined by the method given in the previous Subsection. This equation is derived by assuming that a single use of the decoder will fail with probability $P_{\text{fail}} = A \exp(\alpha d^\beta \log (p / p_{\text{th}}) ) $ in the low-$p$ regime. We then calculate to first order the probability that the decoder will fail a single time in $N+1$ uses to give Eqn.~(\ref{Eqn:LogicalFailureRate}), where we include an additional use of the decoder to account for a possible logical failure during readout.

We manipulate Eqn.~(\ref{Eqn:LogicalFailureRate}) to show a method to evaluate $\alpha$ and $\beta$. We first take the logarithm of both sides of Eqn.~(\ref{Eqn:LogicalFailureRate}) to find the linear expression
\begin{equation}
y = \log((N+1)A) +  \alpha d^\beta u, \label{Eqn:GradientFitting}
\end{equation}
where we write $ y = \log P_\text{fail} $ and $ u = \log (p / p_{\text{th}}) $. We then take the gradient, $g(d) = \dd y / \dd u$, from Eqn.~(\ref{Eqn:GradientFitting}) to find
\begin{equation}
\log g(d) = \log \alpha + \beta \log d. \label{Eqn:FinalFitting}
\end{equation}

\begin{figure}
\includegraphics[width=\columnwidth]{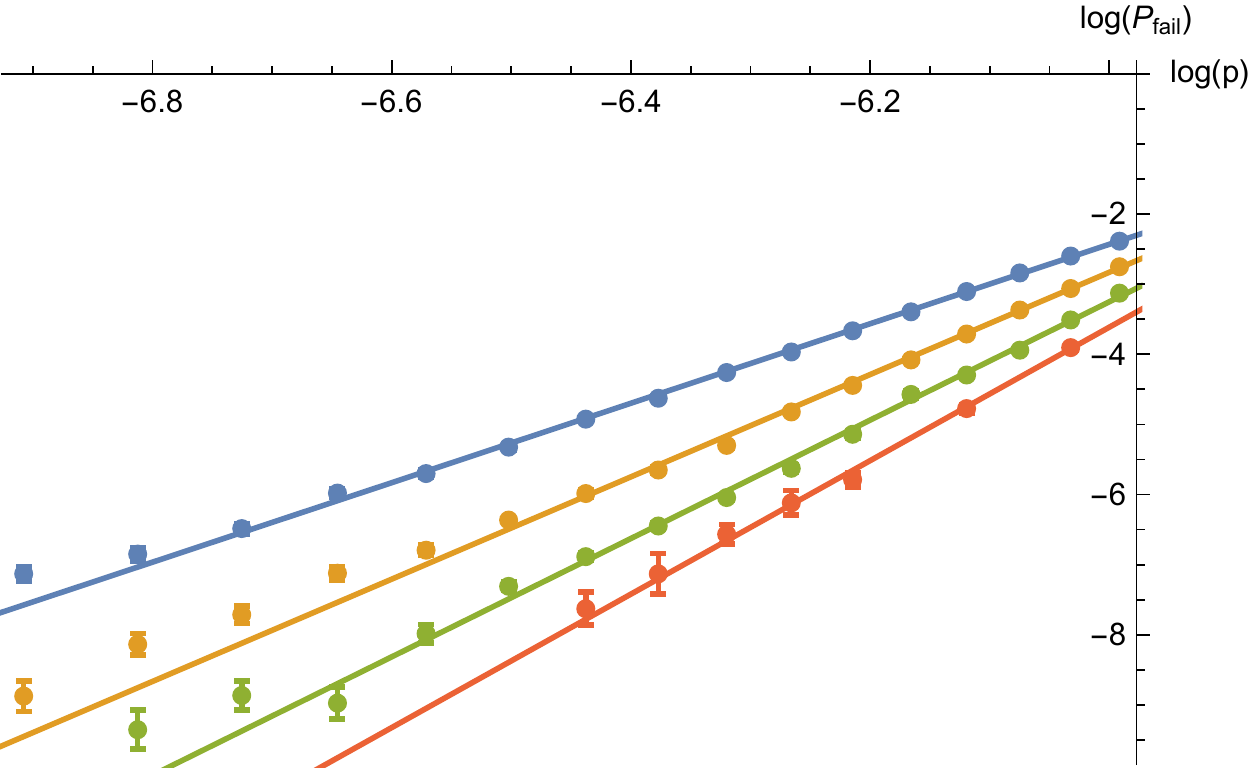}
\caption{\label{Fig:LogPFit}Plot showing $ \log P_{\text{fail}}$ as a function of $ \log p $  for the case that $N = 8$ where we have plotted system sizes $L = 17,\, 23,\, 29 $ and $ 35$ shown in blue, yellow, green and red, respectively. The error bars show the standard error of the mean given by the expression~$\Delta \log p =  \left[ (1 - P_{\text{fail}})P_{\text{fail}} /\eta \right]^{1/2} /  P_{\text{fail}} $ where $\eta$ is the number of Monte Carlo samples we collected. The logarithm of the gradients found for the linear fittings are plotted as a function of $\log d$ in Fig.~\ref{Fig:GradientFitting}. \label{Fig:ExampleData} }
\end{figure}

\begin{figure}
\includegraphics[width=\columnwidth]{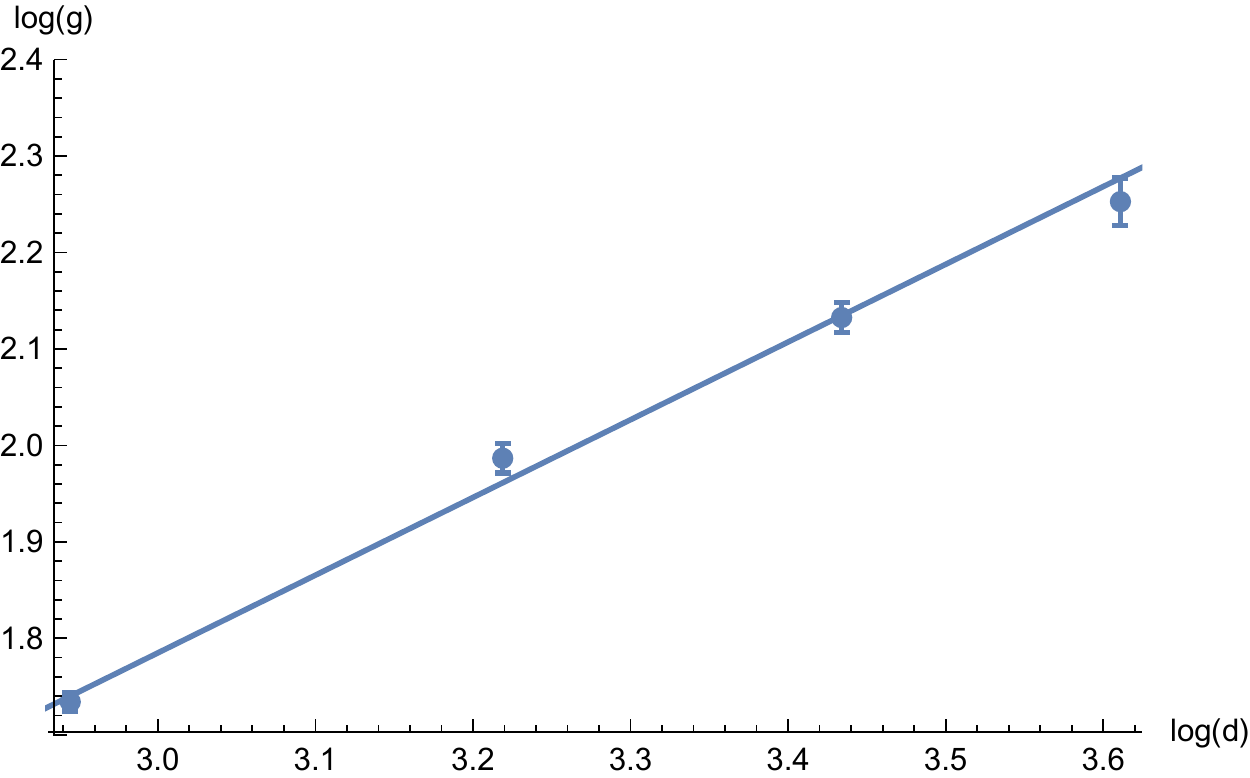}
\caption{Plot shows the available data fitted to the trend anticipated in Eqn.~(\ref{Eqn:FinalFitting}). The Figure shows the logarithm of the gradients found in Fig.~\ref{Fig:LogPFit}, $ \log g(d)$, plotted as a function of the logarithm of the code distance, $\log d$, for the case where $N = 8$. Error bars show the standard error of the mean, evaluated using the {\bf LinearModelFit} function in {\bf Mathematica}. For this case we obtain a fitting $ \log g(d) \approx 0.81 \log d - 0.63  $, as is shown in the plot.  \label{Fig:GradientFitting}}
\end{figure}

In Fig.~\ref{Fig:ExampleData} we plot $\log P_{\text{fail}}$ as a function of $\log p$. For fixed system sizes we observe a linear fitting that Eqn.~(\ref{Eqn:GradientFitting}) predicts. We take the gradient of each of these fittings to estimate $g(d)$, as given in Eqn.~(\ref{Eqn:FinalFitting}). Then, to find $\alpha$ and $\beta$, we plot $\log g(d)$ as a function of $\log d$ where the gradients are taken from the fittings shown in Fig.~\ref{Fig:ExampleData}. This data is shown in Fig.~\ref{Fig:GradientFitting}. We can now determine $ \alpha$ and $\beta$ using, respectively, the $d = 1$ intersection, and the gradient of a linear fit shown in Fig.~\ref{Eqn:GradientFitting}.

The logical failure rates we use to find $\alpha$ and $\beta$ are found using between $10^5$ and $10^6$ Monte Carlo samples for each value of $p$ where we only take values of $p <  0.8 \times p_{\text{th}}$ for each $N$. We discard data points where we observe fewer than ten failures for a given $p$. The data was collected over 5000 CPU hours.

\begin{figure}[t]
\includegraphics[width=\columnwidth]{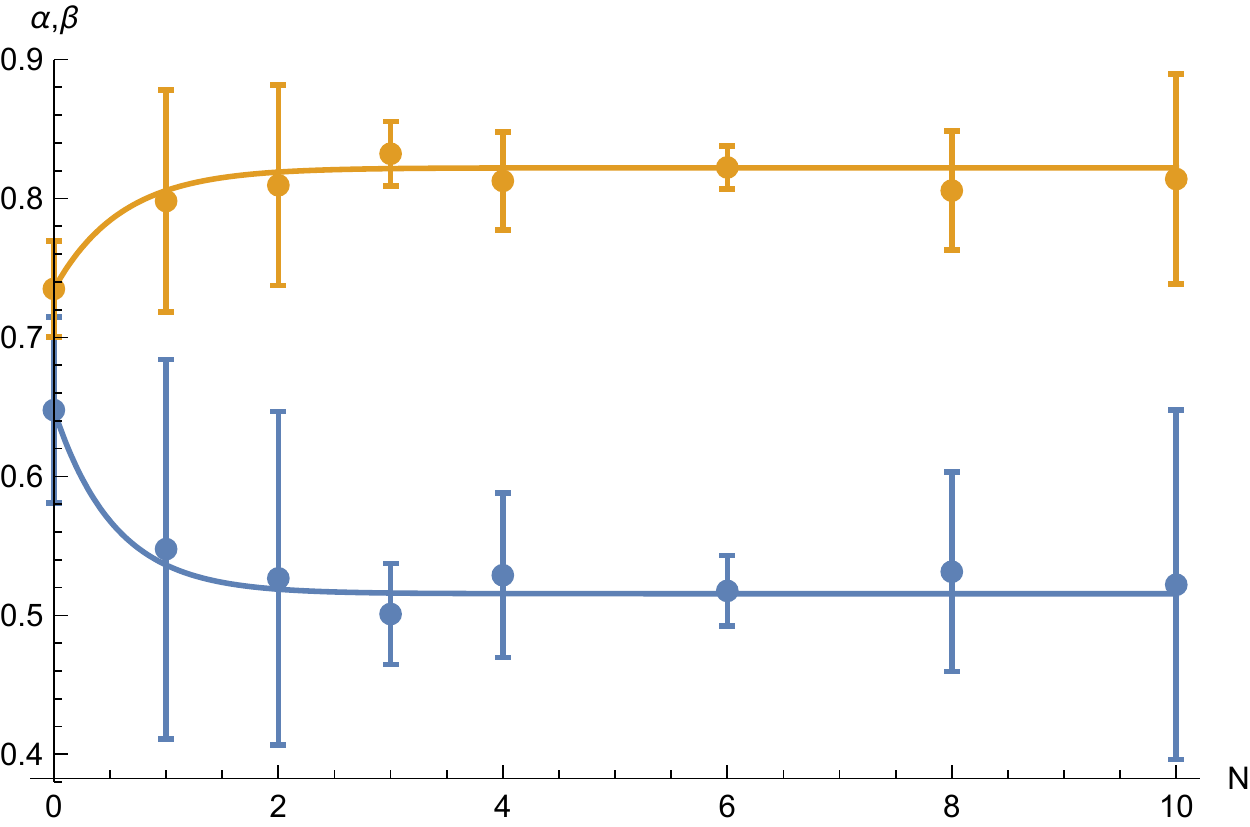}
\caption{\label{Fig:AlphaBetaInN} Unitless values, $\alpha$ and $\beta$, plotted as a function of error-correction cycles, $N$, shown in blue and yellow respectively. The error bars show the standard error of the mean, which are found using the {\bf NonLinearModelFit} function included in {\bf Mathematica}. The $\alpha$ and $\beta$ data points are fitted to Eqns.~(\ref{Eqn:AlphaFitting}) and~(\ref{Eqn:BetaFitting}), respectively. The fitted functions are shown in the plot.}
\end{figure}

We plot the values we find for $\alpha$ and $\beta$ as a function of $N$ in Fig.~\ref{Fig:AlphaBetaInN}. Importantly, we observe convergence in the large $N$ limit. We see this using the fitting functions
\begin{equation}
\alpha(N) = \alpha_\infty \left[ 1 - (1-\alpha_0 / \alpha_\infty) \exp(-\gamma_\alpha N) \right], \label{Eqn:AlphaFitting}
\end{equation}
and
\begin{equation}
\beta(N) = \beta_\infty \left[ 1 - (1-\beta_0 / \beta_\infty) \exp(-\gamma_\beta N) \right], \label{Eqn:BetaFitting}
\end{equation}
to find the $N \rightarrow \infty$ behaviour of our protocol, where $\alpha_0$, $\beta_0$, $\alpha_\infty$, $\beta_\infty$, $\gamma_\alpha$ and $\gamma_\beta$ are constants to be determined, such that
\begin{equation}
\alpha_\infty =  \lim_{N \rightarrow \infty} \alpha(N), \quad \beta_\infty =  \lim_{N \rightarrow \infty} \beta(N). 
\end{equation}
We fit these functions to our data to find 
\begin{equation} 
\alpha_\infty = 0.516 \pm 0.005 , \quad \beta_\infty = 0.822 \pm 0.004 ,
\end{equation}
together with the following values $\alpha_ 0 = 0.65 \pm 0.02$, $\gamma_\alpha = 1.9 \pm 1.5$, $\beta_0 = 0.73\pm 0.01$ and $\gamma_\beta = 1.7\pm 1.4$. The fittings are shown in Fig.~\ref{Fig:AlphaBetaInN}.

\begin{figure}[b]
\includegraphics[width=\columnwidth]{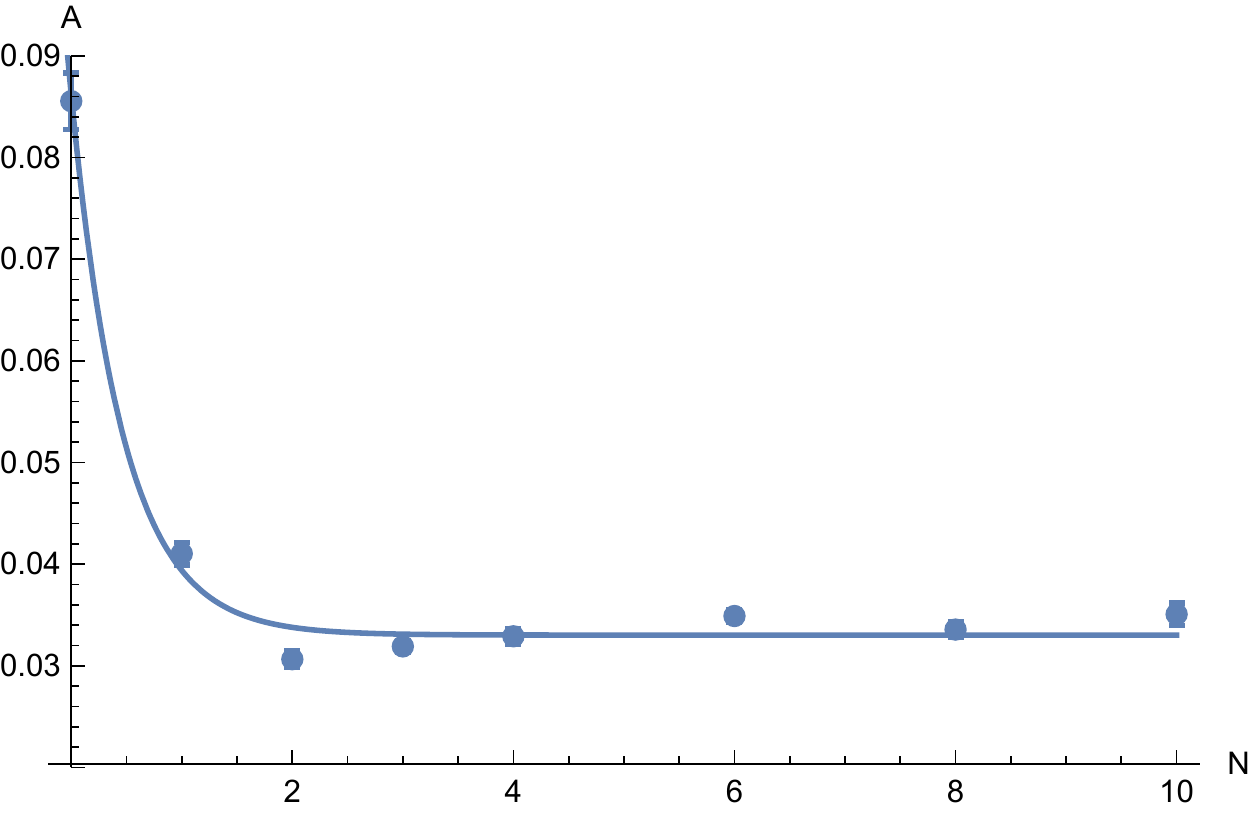}
\caption{\label{Fig:AFitting}Figure shows unitless values of $A$ numerically determined as a function of the number of error-correction cycles we perform, $N$. The error bars show the standard error of the mean which is calculated using {\bf Mathematica}. We fit the data to Eqn.~(\ref{Eqn:AFitting}) as we show in the Figure. }
\end{figure}

Finally, given that we have evaluated $\alpha(N)$ and $\beta(N)$ for different values of $N$, we use these values together with Eqn.~(\ref{Eqn:LogicalFailureRate}) to determine $A$ as a function of $N$. We show the data in Fig.~\ref{Fig:AFitting}. We fit the values of $A(N)$ to the following expression
\begin{equation}
A(N) = A_\infty \left[ 1 - (1-A_0 / A_\infty) \exp(-\gamma_A N) \right], \label{Eqn:AFitting}
\end{equation}
to find values $A_\infty = 0.033 \pm 0.001 $, $A_0 = 0.09 \pm 0.01 $ and $\gamma_A = 2.1\pm 0.5$, thus giving all of the variables, $A_\infty$, $\alpha_\infty$ and $\beta_\infty$, we require to estimate the $N \rightarrow \infty$ behavior of the decoding scheme in the below threshold regime. \\

\bibstyle{plain}

\newpage

\section{Supplementary Figures}

\subsection*{Supplementary Figure 1}
\begin{center}
\includegraphics{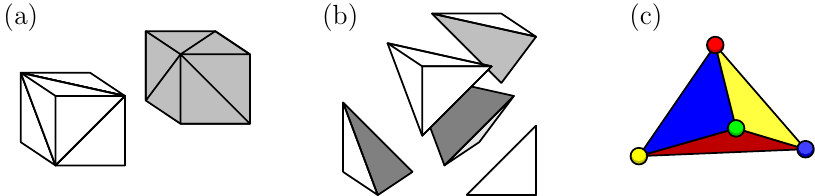}
\end{center}
(a) An odd and an even unit cube that each consist of five tetrahedra. The odd unit cube differs from the even unit cube by a $\pm \pi/2$ rotation about any of the canonical axes. (b) An exploded even unit cell that reveals the internal structure of a unit cube. (c) A fundamental tetrahedron of the lattice whose vertices are correctly four colored. We color each face of every tetrahedron with the color that is not the color of any of its three vertices. The face that is not visible is colored green. \label{FundamentalUnits}

\subsection*{Supplementary Figure 2}
\label{ManyUnitCubes}

\begin{center}
\includegraphics{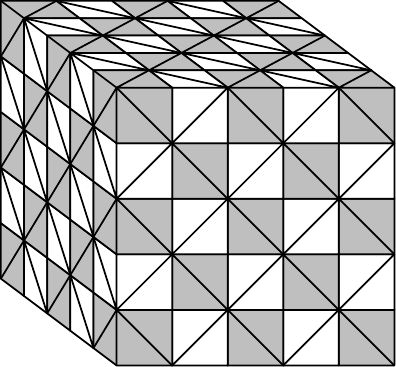}
\end{center}
A cube of linear size $L=5$ formed by stacking odd and even unit cubes. The odd and even unit cubes, shown in grey and white, are stacked such that no two even unit cubes meet at a face and  no two odd unit cubes meet at a face.

\subsection*{Supplementary Figure 3}

\begin{center}
\includegraphics{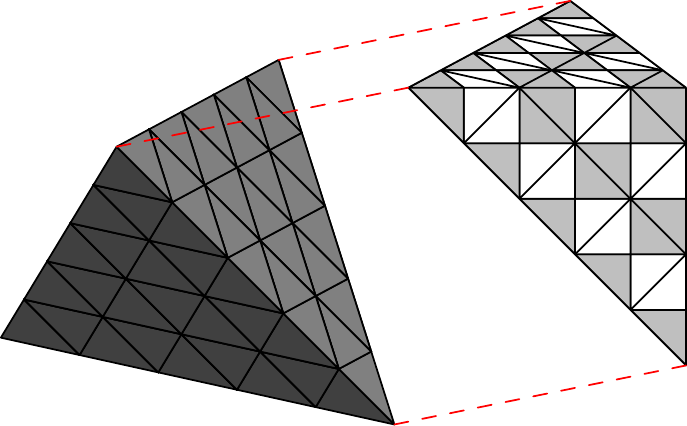}
\end{center}
We remove tetrahedra from the cubic lattice shown in Supplementary Figure 2 to obtain the four-sided tetrahedral structure we require of the gauge color code lattice. The figure shows some of the tetrahedra that have been removed from the cubic lattice to the right of the Figure. The remaining tetrahedra do not define a gauge color code dual lattice without replacing some of the removed tetrahedra, as we explain in Supplementary Note~1 and show in Supplementary Figure~4.

\subsection*{Supplementary Figure 4}

\begin{center}
\includegraphics{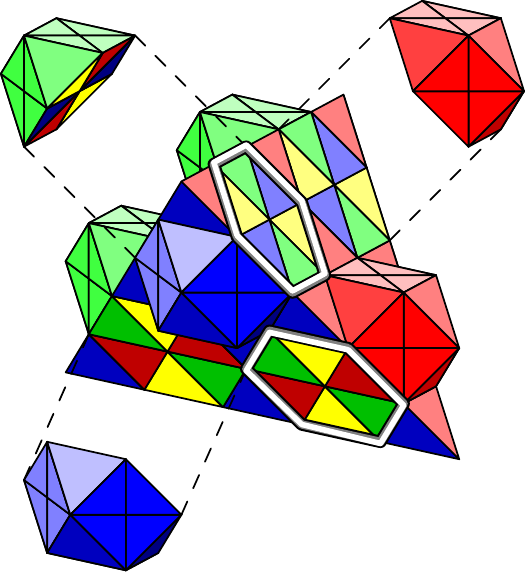}
\end{center}
We modify the lattice of Supplementary Figure 3 to uniformly color the boundaries by replacing some of the tetrahedra that were removed from the cubic lattice. Patches of six incorrectly colored faces are outlined with white hexagons. We add additional tetrahedra to these incorrectly colored patches such that all four boundaries are uniformly colored. The Figure shows some hexagons where the additional tetrahedra are already added. \label{Boundaries}

\subsection*{Supplementary Figure 5}

\begin{center}
\includegraphics{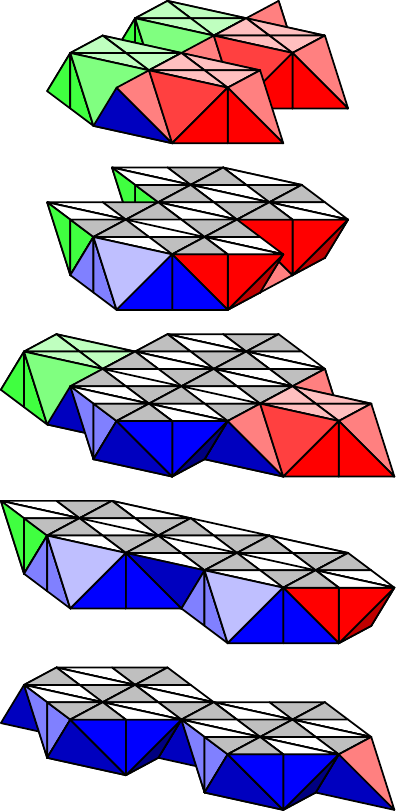}
\end{center}
The $L=5$ lattice where the layers of tetrahedra are separated. Interior tetrahedra are colored white and grey. \label{LatticeLayers}

\subsection*{Supplementary Figure 6}

\begin{center}
\includegraphics{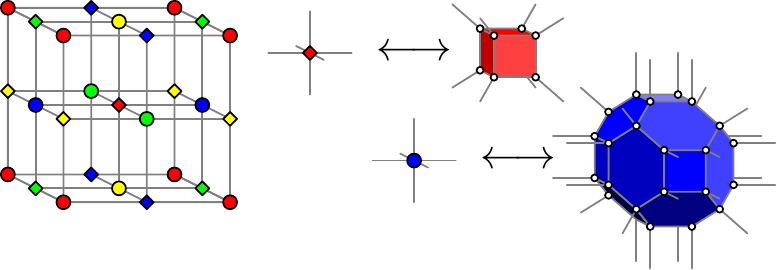}
\end{center}
A repeating cell of the lattice that consists of eight of the fundamental unit cubes. Stabilizers of weight eight and thirty-two are represented by diamonds and circles on the vertices of the repeating cell respectively, as shown in the key at the right of the Figure. Opposite faces of the repeating cell are equivalent. \label{StabilizerGeometry}

\subsection*{Supplementary Figure 7}

\begin{center}
\includegraphics[scale=2.5]{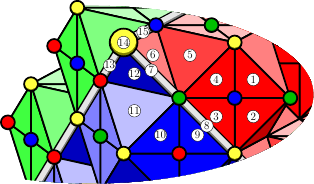}
\end{center}
Qubits on the exterior of the gauge color code lattice. Qubits lie on all the faces, on the edges where two differently colored boundaries meet, fattened and colored in grey in the figure, and on the vertex where three differently colored boundaries meet, marked by a large yellow vertex in the Figure. Qubits are numbered in accordance with the text in Supplementary Note~2. Qubits 7, 8, 13 and 15 lie on lattice edges and qubit 14 lies on the vertex where the red blue and green vertices meet. All other labeled qubits lie on the external faces of the lattice.

\newpage

\subsection*{Supplementary Figure 8}

\begin{center}
\includegraphics{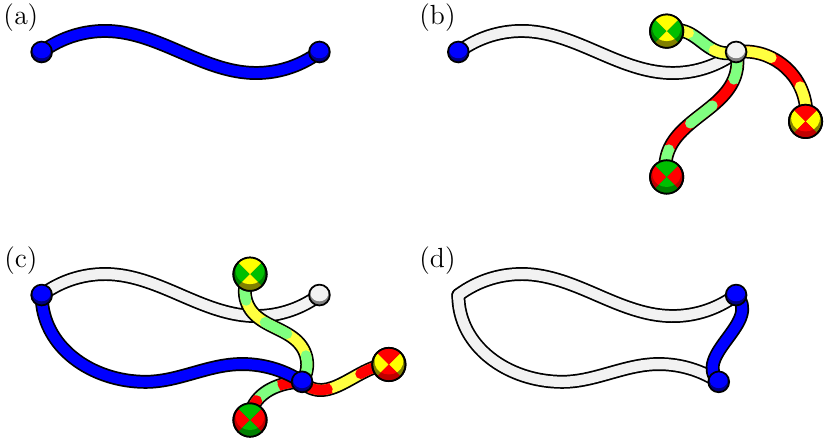}
\end{center}
Errors introduced during fault-tolerant error correction. (a) An error string that, given perfect measurement outcomes, generates two blue stabilizer defects. (b) Measurement errors appear as strings on the dual lattice picture, which are shown as multi-colored strings in the diagram. Gauge defects are identified where strings of measurement errors terminate. The original physical error is marked in grey. (c) During syndrome estimation, we incorrectly estimate the position of the measurement errors, and thus incorrectly identify the true position of the stabilizer defect. The correction operator we apply is represented by a blue string. (d) The correction operator we apply to correct the estimated stabilizer defect introduces a new error to the system. The net error that is introduced by the initial physical error and the inaccurate correction operator effectively acts like the discrepancy between the estimated position of the stabilizer defect from its true position. This discrepancy is marked by the blue error string in the diagram. \label{Fig:CorrelationsDevelop}

\section{Supplementary Notes}

\subsection*{Supplementary Note 1}
\label{SCTN:GCCLattice}

Here we elaborate on the dual lattice we use to construct the gauge color code simulated in the main text. The gauge color code dual lattice is a neatly stacked structure of many tetrahedra. Conveniently though, the lattice dual to the lattice geometry shown in Figure~1(a) in the main text is composed of odd and even cubic units of five tetrahedra, shown in white and grey, respectively, in Supplementary Figure~1(a). We also show the decomposition of an even unit cell in Supplementary Figure~1(b). An odd unit cell differs from an even unit cell only by a $\pi/2$ rotation about any of the canonical axes. Unit cubes on the boundaries of the lattice are modified simply by removing subsets of their tetrahedra.

To ease the explanation of the lattice construction, it is instructive to begin with a cubic block of fundamental cubes, as shown in Supplementary Figure~2. The block must have an odd linear dimension of unit cubes. Tetrahedra are removed from this block to find an appropriate four-sided structure, and finally some of the tetrahedra that have been removed are once again replaced to find suitable boundaries that can be correctly colored.

In the dual representation we require that the vertices of the lattice are four colorable, i.e. we can consistently color every vertex with one of four colors, red, green, yellow or blue, such that no two vertices of the lattice that share an edge can have the same color. We show a single tetrahedra whose vertices are correctly colored in Supplementary Figure~1(c). The vertices of the dual lattice in Supplementary Figure~2 are indeed four colorable.

The gauge color code requires four distinct boundaries where boundaries differ by the subset of colors of the vertices that lie on their surface. Specifically, a boundary of a given color contains no vertices of that color, i.e. a green boundary contains only red, blue and yellow vertices. We therefore look for a tetrahedral structure with four boundaries of four different colors. We design the correct structure by first removing large subsets of fundamental tetrahedra to find the global shape of the lattice. We find a four-sided tetrahedral structure by removing four corners of the cubic block of unit cubes, as shown in Supplementary Figure~3.

The lattice shown in Supplementary Figure~3 does not define the gauge color code. This is because we cannot four-color the lattice such that all four boundaries of the lattice are correctly three colored, as we have explained. However, by replacing some of the tetrahedra we have removed from the lattice, we recover suitable boundaries. To find three-colored boundaries it is convenient to color the faces of the fundamental tetrahedra. We assign to each face of the tetrahedra the color that is not given to any of its adjacent vertices, as shown in Supplementary Figure~1(c). With this face coloring, we require that the lattice has four distinct, uniformly colored boundaries, as shown in the main text in Figure~1(b).

We modify the tetrahedral lattice shown in Supplementary Figure~3 to find a suitable lattice. In Supplementary Figure~4 we color the external faces of the fundamental tetrahedra to show the additional tetrahedra we must replace on the surface of the sheered lattice in Supplementary Figure~3. In particular, we observe hexagonal `patches' of faces which are inconsistently colored compared with the rest of the boundary. We outline some of these hexagonal patches in white. At the centre of each of these hexagons lies a single vertex of a color that is not suitable for the given boundary. To rectify this we replace all of the removed tetrahedra to the lattice that were originally touching the central vertex of each hexagonal patch. Upon doing so we recover the dual lattice shown in the main text. In Supplementary Figure~4 we show some of the `bulbs' of sixteen tetrahedra we reintroduce to the lattice. Some of the bulbs have already been reattached in the Figure. For the convenience of the reader we show the lattice separated into layers in Supplementary Figure~5.

We finally remark that the lattice we consider is particularly convenient for simulation because the stabilizers lie on the vertices of a cubic lattice. In Supplementary Figure~6 we show a repeating unit of the lattice geometry on a cubic lattice. In the Figure stabilizers are represented by circular and diamond-shaped vertices of appropriate color as explained by the key.

\subsection*{Supplementary Note 2}
\label{SCTN:GCCLatticeCharacteristics}

Here we analyse the lattice by counting the total number of qubits as a function of $L$. We also count the number of stabilizers, and the number of gauge operators. We also discuss the qubit support of the stabilizers, and the gauge operators. The quantities evaluated in this Note serve as good sanity checks for readers that are reproducing the gauge color code lattice.

\subsubsection{Qubits}

The gauge color code is a complicated system where qubits are placed on tetrahedra, and on subsets of faces, edges and vertices of the dual lattice we described in Supplementary Note 1. A qubit is placed on each of the exterior vertices where three differently colored boundaries meet. We therefore have vertex qubits 
$$Q_v(L) = 4.$$
 
A qubit is also placed on exterior edges of the lattice where two differently colored boundaries meet. 
We have 
$$ Q_e(L) = 6L,$$
qubits on exterior edges of the lattice. 

The lattice has qubits placed on the exterior faces of the dual lattice. We find that there are
$$Q_f(L) = 9L^2 - 5,$$ 
face qubits. 

Finally, the lattice has a qubit on each of its tetrahedra. We find 
$$Q_t(L) = (5L^3 + 24L^2 - 2L - 24 )/ 3, $$ 
tetrahedron qubits. 

A gauge color code following our construction therefore has $Q(L)= Q_v(L)+Q_e(L)+Q_f(L)+Q_t(L)$ qubits, which is explicitly
$$ Q(L) =  (5L^3 + 51L^2 + 16L - 27)/3, $$
qubits.

\subsubsection{Stabilizers in the Dual Lattice}

Each vertex of the lattice supports two stabilizers. We find the number of vertices
$$ v(L) = (L^3 + 12L^2 +5L - 6)/3. $$
Each stabilizer represented by a given vertex acts on every vertex, edge, face or tetrahedron qubit that contains the respective stabilizer vertex. To make this statement more rigorous, we can specify each object on the lattice that supports a qubit by a list of vertices, $v$, that are contained by the respective object. A vertex qubit is specified by a single vertex, $v = \{v \}$, an edge qubit is specified by a pair of vertices, $e = \{ v_1, v_2  \}$, each triangular face that supports a qubit is specified by three vertices $f = \{ v_1, v_2, v_3 \}$, and a tetrahedron is specified by its four vertices $t = \{ v_1, v_2, v_3, v_4 \}$.

Given that the different objects of the lattice that support qubits, known as simplices, can be uniformly denoted by a lists of vertices of varying length, we are free to group all vertex qubits, edge qubits, face qubits and tetrahedron qubits into the set of qubits, $\mathcal{Q}$, using this simplicial description. Written as simplices, vertex, edge, face and tetrahedron qubits only differ by the length of their list of vertices. Using this notation we can conveniently write down stabilizer operators
\begin{equation}
S^X_v = \prod_{Q\ni v} X_Q,\quad S^Z_v = \prod_{Q\ni v} Z_Q,
\end{equation}
where we take the product of all qubits $Q \in \mathcal{Q}$ that contain vertex $v$. We point out that we have used notation here that is inconsistent with the main text, as here we index stabilizers by vertices, $v$, whereas in the main text we chose the index $c$ to represent cells that support stabilizers. This reflects the change from the primal to dual lattice notation.

\subsubsection{Gauge Operators in the Dual Lattice}

We finally find the number of face operators we use in the simulation in the main text, and explicitly describe the supports of the gauge operators on the dual lattice. Gauge operators are represented on edges, and on exterior vertices of the dual lattice. Instead of counting the number of gauge operators exactly, we count distinct supports of gauge operators. The number of supports are exactly the number of faces on the primal lattice. For each support, $s$, we have two gauge operators, $G^X_s$ and  $G^Z_s$. On the dual lattice, gauge supports are uniquely represented by either an edge, $e$, or a pair $(v,\mathbf{c})$ that contains a vertex, $v$, and a color $\mathbf{c} \in \mathcal{C} \equiv \left\{ \mathbf{r}, \mathbf{g}, \mathbf{y}, \mathbf{b} \right\}$, as is defined in the main text.

To count the gauge supports, we first find the number of edges contained on the lattice, $e(L)$. We find this number using the Euler characteristic 
\begin{equation}
\chi_{\text{3D}}  = v(L) - e(L) + f(L) - t(L), \label{EulerChar}
\end{equation}
where $\chi_{\text{3D}} = 1$ for a `ball-shaped' triangulation, such as that which describes the gauge color code, and where $v(L) $, $f(L)$ and $t(L)$ are the number of vertices, faces and tetrahedra of the lattice, respectively. We already have the number of tetrahedra $t(L) = Q_t(L)$, and $v(L)$ is already counted to find the number of stabilizers. We easily find the number of faces of the lattice $f(L)$ using the fact that each interior face lies on the surface of two tetrahedra, and each tetrahedra has four faces. Following this, up to the exterior faces, we can regard each tetrahedra as contributing half to the face count per face of each tetrahedra. We therefore obtain a contribution of $\sim 2Q_t(L)$ faces for the tetrahedra of the lattice. To account for exterior faces, we must add an additional half-unit per exterior face of the lattice to count the total number of faces, giving the number of faces $f(L) = 2Q_t(L) + Q_f(L)/2$. Explicitly, we have
$$ f(L) =  (20L^3  + 123L^2  - 8L  - 111 )/ 6.  $$
Using $f(L)$ and Eqn.~(\ref{EulerChar}) we find the number of edges
$$ e(L) = ( 4L^3  + 33 L^2  + 2L - 27)/2. $$

Each edge of the lattice represents one gauge support that acts on all edge, face, and tetrahedron qubits that contain the respective edge. We use once again the simplicial notation to denote gauge supports represented by edges, $e$, such that 
\begin{equation}
G^X_{e } = \prod_{Q \ni \{ v_1, v_2 \}} X_Q,\quad G^Z_{e } = \prod_{Q \ni \{ v_1, v_2 \}} Z_Q,
\end{equation}
where we take the product over all qubits $Q \in \mathcal{Q}$ that contain both vertices of edge $e = \{ v_1, v_2 \}$. We point out that no vertex qubit on the lattice can contain the two vertices of an edge, and therefore vertex qubits are not found in gauge operator supports associated to edges.

Gauge supports are also represented by vertices $v$ on the exterior of the lattice. Each exterior vertex represents either one, two or three gauge supports. For exterior vertices that represent multiple gauge operator supports, we specify the supports uniquely with a vertex, $v$, and a color $\mathbf{c}$.

Specifically, a gauge support of a given exterior vertex $v$ contains a {\em subset} of the vertex, edge or face qubits that contain $v$. The subset of lattice objects $Q \ni v$ are taken to be all of those that do not contain any vertices of one particular color, $\mathbf{c}$. 

To make this statement rigorous using the simplex notation, we define the function $C:v \rightarrow \mathcal{C}$ that returns the color of the vertex as defined by the choice of the four-coloring of the lattice. 

External vertex gauge operators are written
\begin{equation}
G_{(v,\mathbf{c})}^X = \prod_{Q \ni v  \backslash \mathbf{c}} X_{Q}, \quad G_{(v,\mathbf{c})}^Z = \prod_{Q\ni v \backslash \mathbf{c}} Z_{Q},
\end{equation}
where we use a shorthand notation `$Q  \ni v  \backslash \mathbf{c}$'  to denote qubits $Q$ that contain vertex $v$, but do not contain any vertices of color $\mathbf{c}$, i.e. where $C(v') \not= \mathbf{c} $ for all vertices $v' \in Q$. We point out that all tetrahedra contain one vertex of each color, and as such, no tetrahedra appear in any gauge supports represented by exterior vertices.

In some cases there are exterior vertices with multiple nonempty subsets $Q  \ni v  \backslash \mathbf{c}$ for different choices of $\mathbf{c}$. In what follows we explicitly consider exterior vertices that contain one, two, and three gauge operator supports. We will see that exterior vertices in the middle of a boundary will contain only one gauge operator support, exterior vertices found where two different colored boundaries meet contain two distinct gauge operator supports, and vertices that lie where three different boundaries meet contain three different gauge operator supports.

Exterior vertices in the middle of a boundary, far away from any edge or vertex qubits, denote one gauge operator support. We explicitly consider the example of the gauge support associated to the blue vertex shown on the red boundary in Supplementary Figure~7. The gauge operator support associated to this vertex act on the qubits labeled 1, 2, 3 and 4. All qubits that contain the exterior vertex of interest contain a yellow, a green, and a blue vertex, but no red vertex.

Vertices that lie at a point where two differently colored boundaries meet represent two distinct gauge operator supports. We consider the green vertex in Supplementary Figure~7 that lies where the blue boundary and the red boundary meet. One of the supports acts on the qubits labeled 3, 4, 5, 6, 7 and 8. Common to all of these qubits is that none of their faces or edges contain a red vertex. The other support acts on qubits 7, 8, 9, 10, 11 and 12. Different from the first support we discussed that is associated to the green vertex, none of the faces or edges involved in this support contain a blue vertex.

We finally consider the support of the gauge operators associated to the vertices that lie where three boundaries meet, such as at the large yellow vertex shown in Supplementary Figure~7 that contains the qubit indexed 14. This vertex denotes three supports, two of which are shown in the diagram. One support acts on qubits 6, 7, 14 and 15, the qubits that contain the yellow vertex of interest, but do not contain a red vertex. The other support acts on the qubits numbered 7, 12, 13 and 14, the lattice objects that contain the large yellow vertex, but do not contain any blue vertices. The third support associated to the large yellow vertex acts on the green face qubit that contains the large yellow vertex which cannot be seen in the diagram, together with qubits 13, 14 and 15.

We can now count the number of gauge supports we must measure to realize fault-tolerant error correction using the gauge color code. We count the number of exterior vertices $v^E(L)$ with the expression
 $$ v^E(L) = (9L^2  - 1)/2 . $$
Then, using that the $6(L-1)$ vertices that lie where two boundaries meet represent two gauge operator supports, and four vertices that lie where three boundaries meet represent three gauge operator supports, we find the number of gauge operator supports $G(L) = e(L) + v^E(L)  + 6(L-1) + 8$, giving the number of gauge operator supports
$$
G(L) = 2L^3 + 21L^2 + 7L - 12.
$$
Twice this quantity gives the number of face operators we use to perform fault-tolerant error correction with the gauge color code to identify both Pauli-X and Pauli-Z type errors.

\subsection*{Supplementary Note 3}
\label{SCTN:CorrelatedErrorsDevelop}

In the main text we identify the threshold error rate as a function of the number of rounds of error correction that are performed, $N$, to show that the threshold error rate is robust in the limit $N\rightarrow \infty$. It is important to look at the single-shot error-correction scheme after repeated applications because, when one considers noisy measurements, the error-correction protocol will leave some residual noise on the code. In general, this residual noise is not easily characterized. In fact, the noise introduced to the system is a complex function of the physical error rate, the measurement error rate, and the choice of error-correction protocol. Given that the nature of the residual noise is not well understood, it is not clear that the error-correction protocol will be able to successfully deal with the residual noise after many cycles of error correction.

In Supplementary Figure~8 we give an example of a mechanism that leads to the development of a correlated error. We show a physical error in Supplementary Figure~8(a). The error is represented by a string. Given perfect measurements, we expect to observe two stabilizer defects at the left and right ends of the error string. 

For this example, some measurement errors occur in the vicinity of the right stabilizer defect when we attempted to learn the positions of stabilizer defects. The measurement errors, where face operators return incorrect measurement outcomes, are represented by multi-colored strings in Supplementary Figure~8(b). Gauge defects lie at the points where the strings of measurement errors terminate. Given these measurement errors occur, we cannot be sure of the true location of the right stabilizer defect. Instead, we can make use of the positions of the three gauge defects, and knowledge of the noise model, to attempt to determine the location of the stabilizer defect.

Our ability to predict the positions of stabilizer defects depends on our choice of syndrome estimation algorithm. In Supplementary Figure~8(c) we show where the syndrome estimation algorithm mistakenly predicts measurement errors. The measurement errors that have been predicted are represented by strings that terminate at the gauge defects, and branch at the estimated stabilizer defect. The incorrect estimation made by the algorithm leads us to believe that the stabilizer defect is not in its true location, but is displaced to some estimated location. 

Next, we apply a correction operator to attempt to repair the initial physical error drawn in Supplementary Figure~8(a). However, using the estimated position of the right stabilizer defect, we apply a correction operator that connects the left stabilizer defect to the estimated right stabilizer defect, as shown by a blue string in Supplementary Figure~8(c), thus introducing additional errors to the code. The initial error, and the applied correction operator is shown in grey in Supplementary Figure~8(d). 

Due to the topological nature of the string errors in the gauge color code, the net effect of the initial error and the correction operator equates to the discrepancy between the true position of the right stabilizer defect, and the position of the estimated stabilizer defect. The effective error is shown by the blue error string in Supplementary Figure~8(d). This is easily checked as we can continuously deform the grey string onto the blue string. In other words, in the gauge color code the grey error string is equivalent to the blue error string up to multiplication by gauge operators.

In general it is not clear how able a syndrome estimation algorithm is for correctly predicting the positions of stabilizer defects. A bad syndrome estimation algorithm might displace stabilizer defects over long distances compared to their true positions, in which case large correlated errors that we cannot correct may develop. Moreover, it is not clear that the character of the residual noise remains constant over many error-correction cycles. Indeed, one should be concerned that an error correction protocol might cause correlations to develop over repeated use of the error-correction procedure while information is stored. It is therefore important to study a single-shot error-correction protocol over many cycles of error correction to interrogate its performance, and to check that the noise incident to a code achieves a steady state in the long-time limit. In doing so, we are able to establish that a code has a finite threshold after an arbitrarily long time.

\end{document}